\documentclass[singlecolumn,prl,draft]{revtex4}
\usepackage[T1]{fontenc}
\usepackage[utf8]{inputenc}
\usepackage{graphicx}
\usepackage{amsfonts}
\usepackage{amsmath,amsthm,amssymb}
\usepackage[hidelinks]{hyperref}

\begin{document}

\title{Comment on "AdS nonlinear instability: moving beyond spherical symmetry"}

\author{Andrzej Rostworowski}

\email{arostwor@th.if.uj.edu.pl}
\affiliation{M. Smoluchowski Institute of Physics, Jagiellonian University, 30-348 Krak\'ow, Poland}

%\date{\today}

\maketitle
%%%%%%%%%%%%%%%%%%%%%%%%%%%%%%%%%%%%%%%%%%%%%%%%%%%%%%%%%%%%%%%%%%%%%%%%%%%%%%
%%%%%%%%%%%%%%%%%%%%%%%%%%%%%%%%%%%%%%%%%%%%%%%%%%%%%%%%%%%%%%%%%%%%%%%%%%%%%%
%%%%%%%%%%%%%%%%%%%%%%%%%%%%%%%%%%%%%%%%%%%%%%%%%%%%%%%%%%%%%%%%%%%%%%%%%%%%%%

In  a recent paper \cite{ds}, Dias and Santos considered gravitational perturbations of four dimensional anti-de Sitter spacetime and argued that only in special cases (listed in Sec.6 in \cite{ds}) linear eigenmodes admit a nonlinear extension to a regular time-periodic solution  (geon). The reason is that at the third order of the perturbation expansion around most eigenmodes there appear resonant terms that, in contrast to spherically symmetric scalar perturbations studied in \cite{br,mr}, cannot be removed by a frequency correction.  The aim of this comment is to point out that this is a purely technical obstruction in constructing geons, due to the degeneracy of the spectrum and, when this degeneracy is properly taken into account, one can construct perturbatively a geon bifurcating from \emph{each} linear eigenfrequency.

To illustrate our point, we restrict to the simplest case of axially symmetric scalar-type perturbations, for which the eigenfrequencies are (we set the AdS radius to one)
\begin{equation}\label{spectrum}
\omega_{\ell,p}=1+\ell+2p,
\end{equation}
where $\ell\geq 2$ is the angular momentum index and a nonnegative integer $p$  is the nodal number for radial mode functions
\begin{equation}\label{radial}
e_{\ell,p}(r) = \frac{r^{\ell + 1}} {\left(r^2+1\right)^{\frac{\ell + 1}{2}}} \, {}_2F_1\left(-p,1+\ell+p;\frac{1}{2};\frac{1}{r^2+1}\right).
\end{equation}
It follows from \eqref{spectrum} that (restricting to axial symmetry and to scalar-type perturbations only) all eigenfrequencies but $\omega_{2,0}=3$ and $\omega_{3,0}=4$ are degenerate. According to the first three rows in Table~1 in \cite{ds}, the eigenmode with the single eigenfrequency $\omega_{2,0}=3$ (the first row in Table~1) can be extended to a geon (the same happens for the eigenmode with frequency $\omega_{3,0}=4$). To see that the double eigenfrequency $\omega=5$ (listed in the second and third row in Table~1) also gives rise to a geon, consider a master gauge invariant variable for scalar-type perturbations 
\begin{equation}
%\label{}
\Phi(t,r,\theta) = \epsilon \, \Phi^{(1)}(t,r,\theta) + \epsilon^2 \, \Phi^{(2)}(t,r,\theta) +  \epsilon^3 \, \Phi^{(3)}(t,r,\theta) + \mathcal{O}\left( \epsilon^4 \right) 
\end{equation}
with a linear combination of two  eigenmodes with $\omega=5$ as the seed:
\begin{equation}
\label{seed}
\Phi^{(1)}(t,r,\theta) = \left(\eta \, e_{2,1}(r) P_2(\cos \theta) + (1-\eta) \, e_{4,0}(r) P_4(\cos \theta) \right) \cos\left((5 + \epsilon^2 \omega_2)t \right) \, ,
\end{equation}
where $P_{\ell}$ are Legendre polynomials. At the third order we get two resonant terms (for the modes $e_{2,1}$ and $e_{4,0}$) that can be removed by a suitable choice of the frequency correction $\omega_2$ and the mixing parameter $\eta$ in \eqref{seed}. More precisely, the resonant terms will be absent if $\omega_2$ and $\eta$ satisfy the following system
\begin{align}
\label{eq:l2}
-651980329\, \eta^3 +673396185\,\eta^2 -358711575\,\eta + 22494375 &= 49201152 \, \eta \,\omega_2
\\
\label{eq:l4}
16847182891\, \eta^3 - 38330631185\, \eta^2 + 31825994625\, \eta - 10200766875 &= 4182097920 \, (1-\eta ) \,\omega_2 \,.
\end{align}
This system has two real solutions: $(\eta,  \omega_2) \approx (0.11433,\,-1.9001)$ and $(\eta, \omega_2) \approx (1.0066, \,-6.4870)$, thus we expect two geons to bifurcate from the eigenfrequency $\omega=5$. Note that setting $\eta=1$  in (\ref{eq:l2}), we get $\omega_2 = -34397/5376$, while setting $\eta=0$  in (\ref{eq:l4}), we get $\omega_2 = -52311625/21446656$,  in agreement  with the values given by Dias and Santos in Table~1. 

In general, for the eigenfrequency with geometric multiplicity $k$ we need to take a linear combination of $k$ corresponding eigenmodes as the seed,  and  to remove the resonances at the third order we have to fulfill the system of $k$ equations that are cubic in the mixing parameters $\eta_1,..,\eta_{k-1}$ and linear in  $\omega_2 \eta_i$ terms (with $1 \leq i \leq k-1$). Each \textit{real} root of this system gives rise to a geon bifurcating from the given eigenfrequency (the number of all (complex) roots is expected to grow exponentially with $k$, as after eliminating $\omega_2$, we are left with the system of $k-1$ equations that are quartic in $\eta_1,..,\eta_{k-1}$). Interestingly, in all cases that we have studied so far (in axial symmetry, both  for  scalar and vector seeds), the number of bifurcating time-periodic solutions is \textit{equal} to the multiplicity of the eigenfrequency. This intriguing coincidence deserves further studies. Finally, we remark that in a parallel work Maliborski constructed axially symmetric time-periodic solutions for the cubic wave equation on the fixed AdS background \cite{m}, and in that model there is no such coincidence.

\noindent
\subsubsection*{Acknowledgements.} This research was supported in part by the Polish National Science Centre grant no. DEC-2012/06/A/ST2/00397. The author is indebted to Piotr Bizo\'n and Maciej Maliborski for many valuable discussions. The author acknowledges the six months \textit{scientific associate} contract at CERN TH department that allowed him for the detailed studies of gravitational perturbations of AdS, and is particularly grateful to Luis \'Alvarez-Gaum\'e for providing propitious atmosphere for the research.

%%%%%%%%%%%%%%%%%%%%%%%%%%%%%

\end{document}